\documentclass[12pt]{article}

\date{}
\usepackage{amsmath,amsfonts,graphicx}
\begin{document}

\title {\bf Topological Quantum Field via Chern-Simons Theory, part1}
\author {Ke Wu, Yifan Zhang \\
Department of Mathematics, Capital Normal University,\\
Beijing, 100048, P.R.China}
 \maketitle

\begin{abstract}
To understand what does Chern-Simons with compact Lie group(does not
like Dijkgraaf-Witten model with finite group in 3d) attach to a
point, we first give a construction of Topological Quantum Field
Theory(TQFT) via Chern-Simons theory in this paper. We discuss the
Topological Quantum Field Theory and Chern-Simons theory via
Category, then interpret the cobordism as cospan and field of
space-time as span, which ultimately deduce the construction of
TQFT.

\end{abstract}

\section{Introduction}

In this paper, we describe a construction for an Topological Quantum
Field Theory via Chern-Simons theory, based on Freed's works which
view topological field as a general set of axioms [F1][F2][F3][F4]
and Morton's work which gives a construction for an Extended
Topological Quantum Field Theory(ETQFT), associating to any finite
gauge group [M]. We can find that when we say the functor
\begin{equation}
Z: nCob \longrightarrow  Hilb
\end{equation}
is monoidal\ (i.e. is a $Hilb$-valued TQFT), it is equivalent to say
that the Chern-Simons field satisfies the "additivity law" which is
the
important property of a lagrangian field theory.\\
\indent
  In the axiomatic formulation (due to Atiyah [At]), an $n$-dimensional
topological quantum field theory is a rule $\Lambda$ which to each
closed oriented manifold $\Sigma$ (of dimension $(n-1)$)associates a
vector space $\Lambda \Sigma $, and to each oriented $n$-manifold
$M$ whose boundary is $\Sigma$ associates a vector in $\Lambda
\Sigma $. The axioms amount to saying that a TQFT is a (symmetric)
monoidal functor from $nCob$ to $Vect_k$, i.e.
\begin{equation}
Z: nCob \longrightarrow Vect_k
\end{equation}
where $nCob$ is the geometric category whose objects are closed
$(n-1)$-dimensional oriented manifolds and morphisms are
$n$-dimensional oriented manifolds as cobordisms of
$(n-1)$-dimensional closed oriented manifolds, and $Vect_k$ is the
algebraic category whose objects are vector spaces over $k$ and
morphisms are $k$-linear maps between vector spaces. I will give the
concrete definition and the example of $2d$-TQFT in $\S2$\ [Ab].
Mathematical interest in $TQFTs$ stems from the observation that
they produce invariants of closed manifolds\ [S]. Meanwhile,
Physical interest in $TQFTs $ comes mainly from the observation that
$TQFTs $ possess certain features one expects from a theory of
quantum gravity and approach to quantum field theory\ [DW]. It
serves as a model in which one can do calculations and gain
experience before embarking on the
quest for full-fledged theory.\\
\indent
  The "degroupoidification" functor gives a representation of $Span(Gpd)$ in
$Veck_k$ which can views as a TQFT that be constructed via gauge
theory associates to any finite gauge group. The reader may find
more details on this program in a review by Baez, Hoffnung and
Walker [BHW], though it is not obvious to refer. In this paper, we
get a construction via Chern-Simons theory associated to any
compact Lie group.\\
\indent
  For a compact Lie group (usually restrict to connected and simply
connected case), and integral cohomology class $\lambda \in
H^{4}(BG)$, we can construct Chern-Simons theory on $3d$ manifold\
[F2][SW]. If $X$ is a closed, oriented $3d$-manifold, then the
action is a complex number with unit norm. This is the exponential
of $2\pi i$ times the usual action, which is a real number
determined modulo the integers. If $X$ has a nonempty boundary, then
the action is an element of unit normal in an abstract metrized
complex line which depends on the restriction of the field to
$\partial X$. This line is called Chern-Simons line, and we can
prove the action satisfies the "Functoriality" "Orientation"
"Additivity" and "gluing", which illustrate the important properties
of a lagrangian field theory. This part
will appear in $\S3$. \\
\indent
   To construct the TQFT, we first should construct a functor $\natural$
from $nCob$ to $Span(FT)$ of field space category. In $\S4$, we view
cobordisms as cospan and field space as span. Then the composition
of span is just the fiber product of groupoid, which will explain
the "additivity law" of the Chern-Simons action.\\
\indent
   The most important part of this article is $\S5$, which gives us
the construction of TQFT via the functor $\natural$. For
Chern-Simons theory, we can get a Hilbert space from any oriented
closed $2d$-manifold and a linear map from the span views as the
cobordism via the functor $\natural$. That is to say we get another
functor $\heartsuit$
\begin{equation}
\heartsuit : {Span(FT)} \longrightarrow {Hilb}
\end{equation}
Then we compose the two functors to get the TQFT.

\section{Topological Quantum Field Theories}
\indent
   Consider the fundamental formula of quantum field
theory
\begin{equation}
\langle{\hat{A_2}|U|\hat{A_1}\rangle} =
\int_{\Sigma_1=A_1}^{\Sigma_2=A_2}{\cal D}A \ exp\{iS[A]\}
\end{equation}\\
\indent
 We have now developed the appropriate mathematical language
to understand this equation. We thus define a topological quantum
field theory ($Hilb$-valued TQFT) as a symmetric monoidal functor
\begin{equation}
Z: nCob \longrightarrow Hilb \nonumber
\end{equation} \\
\indent
   To begin, we must give some explicit information about the
categories of $nCob$ and $Hilb$ we are interested in.\\

\subsection{The Category $nCob$}

\indent
   The category which quantum field theory concerns itself with
is called $nCob$, the "n dimensional cobordism category" [B], and
the rough definition is as follows. Objects are oriented closed
(that is, compact and without boundary) $(n-1)$-manifolds $\Sigma$,
and arrows $M : \sigma_1 \longrightarrow \sigma_2$ are compact
oriented $n$-dimensional manifold $M$ which are cobordisms from
$\Sigma_1$ to $\Sigma_2$. Composition of cobordisms $M : \sigma_1
\longrightarrow \sigma_2$ and $N : \sigma_2 \longrightarrow
\sigma_3$ is defined by
gluing $M$ to $N$ along $\Sigma_2$.\\
\indent
   Let us fill in the details of this definition. Let $M$ be an
oriented $n$-manifold with boundary $\partial M$. Then one assigns
an induced orientation to the connected components $\Sigma$ of
$\partial M$ by the following procedure. For $x\in \Sigma$, let
$(v_1,\cdots,v_{(n-1)},v_n)$ be a positive basis for $T_{x}M$ chosen
in such a way that $(v_1,\cdots ,v_{(n-1)})\in T_x\Sigma$. It makes
sense to ask whether $v_n$ points inward or outward from $M$. If it
points inward, then an orientation for $\Sigma$ is defined by
specifying that $(v_1,\cdot, v_{(n-1)})$ is a positive basis for
$T_{x}M$. If $M$ is one dimensional, then $x \in \partial M$ is
defined to have positive orientation if a positive vector in
$T_{x}M$ points into $M$, otherwise it is defined to have negative
orientation.\\
\indent
   Let $\Sigma_1$ and $\Sigma_2$ be closed oriented $(n-1)$-manifolds.
An cobordism from $\Sigma_1$ to $\Sigma_2$ is a compact oriented
$n$-manifold $M$ together with smooth maps
\begin{equation}
\Sigma_1 \longrightarrow M \longleftarrow \Sigma_2 \nonumber
\end{equation}
where $i$ is a orientation preserving diffeomorphism of $\Sigma$
onto $i(\Sigma_1)\in M$, $i^{'}$ is an orientation reversing
diffeomorphism of $\Sigma^{'}$ onto $i^{'}(\Sigma_2)\in M$, such
that $i(\Sigma_1)$ and $i^{'}(\Sigma_2)$ (called the in- and
out-boundaries respectively) are disjoint and exhaust $\partial M$.
Observe that the empty set $\o$ can be considered as an
$(n-1)$-manifold.\\
\indent
   For a given $n$, one can construct the (smooth) cobordism
category $nCob$. Objects are closed, oriented $(n-1)$-manifolds
$\Sigma$, and arrows $M : \sigma_1 \longrightarrow \sigma_2$ are
cobordism. In order to make this a well-defined category with
identity arrows, we must quotient out diffeomorphic cobordisms.
Specifically, let $M$ and $M^{'}$ be cobordisms from $\Sigma_1$ to
$\Sigma_2$. Then they are considered equivalent if there is an orientation
preserving diffeomorphism $\psi : M \longrightarrow M^{'}$ making
the diagram commute.\\
\indent
    After identifying equivalent cobordisms, the "cylinder" $\Sigma
\times [0,1]$ functions as the identity arrow for $\Sigma$. The
cobordism category is a geometric category which captures the way
$n$-manifolds glue together.\\
\indent
    Furthermore, $nCob$ is a symmetric monoidal category with the
disjoint union as its monoidal product, $\o$ as its identity, and
twist diffeomorphism $T_{\Sigma_1,\Sigma_2} :\Sigma_{1} \sqcup
\Sigma_{2} \longrightarrow \Sigma_{2} \sqcup \Sigma_{1}$ satisfies
the symmetric condition. We denote this symmetric monoidal category
$(nCob,\sqcup,\o,T)$. For more detail, you can read the book [K].

\subsection{The Category $Hilb$}
\indent
   The category $Hilb$ is simple whose objects are Hilbert spaces over $k$,
and arrows are bounded linear maps between Hilbert spaces.
Obviously, $Hilb$ is a symmetric monoidal category which we usually
denote it $(Hilb,\otimes,k,\sigma)$ where $\otimes$ is the common
tensor product of vector space and symmetric condition is hold for
$\sigma :A\otimes B \longrightarrow B\otimes A$. We don't focus much
energy on this category.

\subsection{The definition of TQFT}
\indent
    {\bf Definition 2.3.1} An $n$-dimensional topological quantum field theory (TQFT) is a
rule $\Lambda$ which to each closed oriented $(n-1)$-manifold
$\Sigma$ associates a vector space $\Lambda \Sigma$, and to each
oriented cobordism $M: \Sigma_1 \longrightarrow \Sigma_2$ associates
a linear map $\Lambda M$ from $\Lambda \Sigma_1$ to $\Lambda
\Sigma_2$. This rule $\Lambda$ must satisfy the following five
axioms.
\begin{enumerate}
\item Two equivalent cobordisms must have the same image:\\
\begin{equation}
M \cong M^{'} \Longrightarrow \Lambda M = \Lambda M^{'} \nonumber
\end{equation}
\item The cylinder $\Sigma \times [0,1]$, thought of as a cobordism
from $\Sigma$ to itself, must be sent to the identity map of
$\Lambda \Sigma$.
\item Given a decomposition $M=M_{1}M_{2}$ then\\
\begin{equation}
\Lambda M=(\Lambda M_1)(\Lambda M_2)\qquad ({\rm composition \ of \
linear \ maps}) \nonumber
\end{equation}
\item Disjoint union goes to tensor product: if
$\Sigma=\Sigma_{1}\sqcup \Sigma_2$ then $\Lambda \Sigma =\Lambda
\Sigma_{1} \otimes \Lambda \Sigma_2$. This must also hold for
cobordisms: if $M:\Sigma_0 \longrightarrow \Sigma_1$ is the disjoint
union of $M^{'}:\Sigma_0^{'} \longrightarrow \Sigma_1^{'}$ and
$M^{''}:\Sigma_0^{''} \longrightarrow \Sigma_1^{''}$ then $\Lambda
M=\Lambda M^{'} \otimes \Lambda M^{''}$.
\item The empty manifold $\Sigma=\o$ must be sent to the ground
field $k$.
\end{enumerate}
\indent
    {\bf Remark :} The first two axioms express that the theory is topological: the
evolution depends only on the diffeomorphism class of space-time,
not on metric structure. Axiom (4) reflects a standard principle of
quantum mechanics: that the state space of two independent systems
is the tensor product of the two state spaces.\\
\indent
   Towards a categorical interpretation of the axioms of the
topological quantum field theory [B], we say a $Hilb$-valued TQFT
[L]is just a symmetric monoidal functor from $(nCob,\sqcup,\o,T)$ to
$(Hilb,\otimes,k,\sigma)$, i.e.
\begin{equation}
Z : (nCob,\sqcup,\o,T) \longrightarrow (Hilb,\otimes,k,\sigma)
\end{equation}

\subsection{$2d$-topological quantum field theory}
\indent
    From the mathematical view, the classification of $nd-TQFT$ is
very important. For general $n$, this question is so difficult to
solve. But the classification of Extended TQFT has been complete
done by Lurie [L]. In this subsection, we will outline the results of the
classification of $2d$-TQFT, rather than one of Extended TQFT.\\
\indent
    The first statement that $2d-TQFTs$ are commutative Frobenius
algebras appears in Dijkgraaf's thesis (89) [D], but mathematical
proofs didn't appear until the work of L.Abrams(95), S.Swain(95),
T.Quinn(95), and B.Dubrovin(96). In order to give the
correspondence, we first need the following "generators-relations"
theory\ [K]:\\
\indent {\bf Theorem 2.4.1} The monoidal category $2Cob$ is generated
under
composition and disjoint union by the following six cobordisms:\\

\newpage

\begin{figure}[h]
  \centering
  \includegraphics[width=1\textwidth]{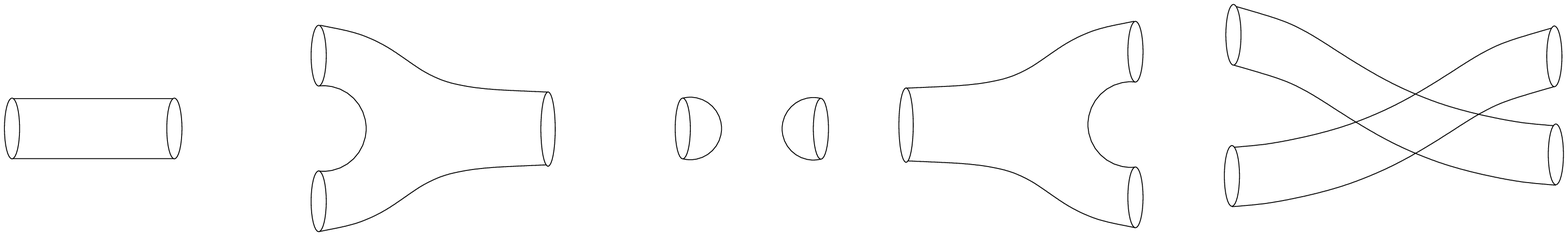}
\end{figure}

and, those cobordisms satisfy the following relations (figure:1-6).\\

\begin{figure}[!h]
  \centering
  \includegraphics[width=1\textwidth]{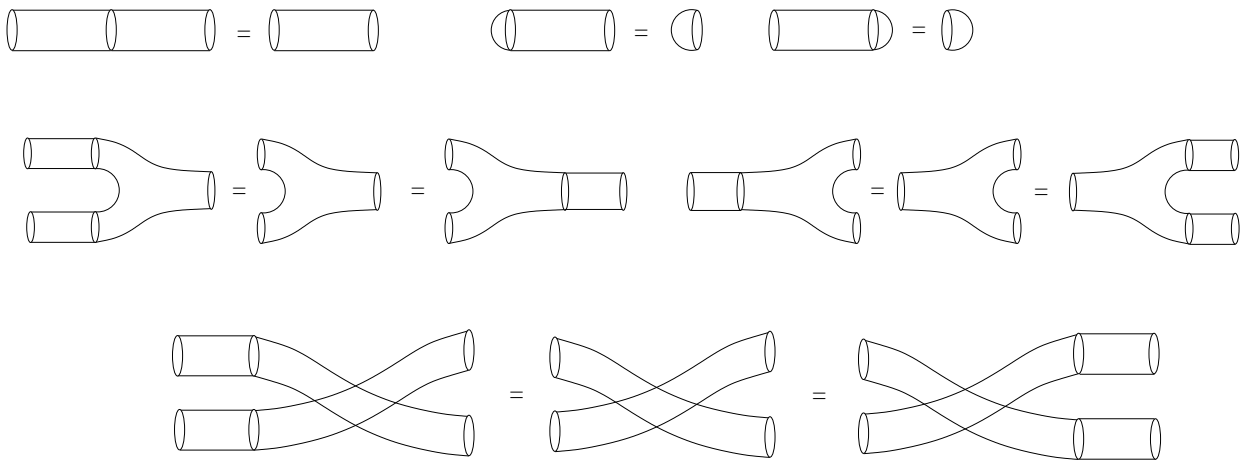}
  \caption{identity relations}
\end{figure}

\begin{figure}[!h]
  \centering
  \includegraphics[width=1\textwidth]{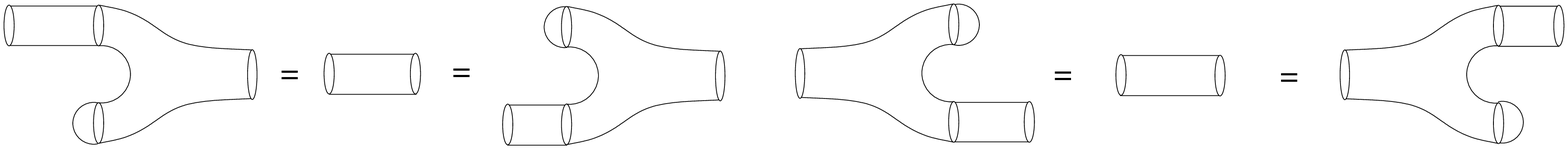}
  \caption{sewing indices}
\end{figure}

\begin{figure}[!h]
  \centering
  \includegraphics[width=1\textwidth]{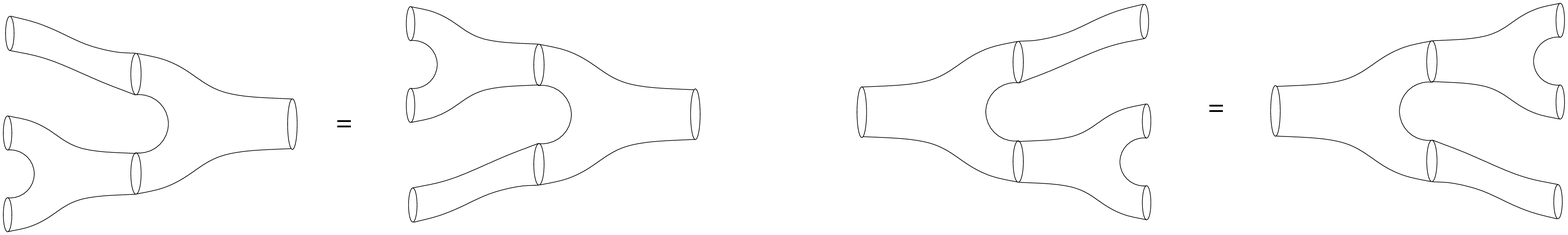}
  \caption{associativity and coassociativity}
\end{figure}

\begin{figure}[!h]
  \centering
  \includegraphics[width=1\textwidth]{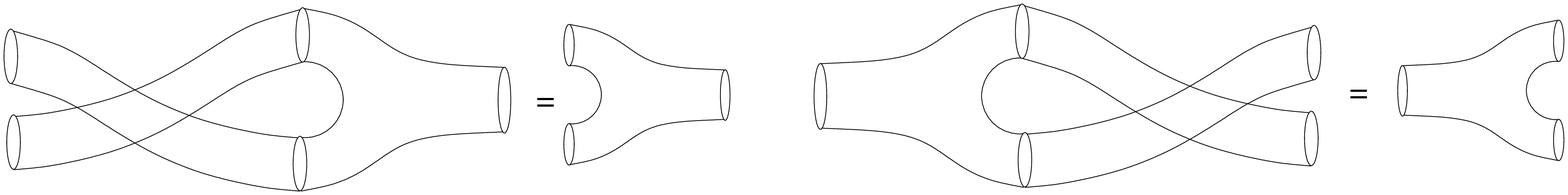}
  \caption{commutativity and cocommutativity}
\end{figure}

\begin{figure}[!h]
  \centering
  \includegraphics[width=1\textwidth]{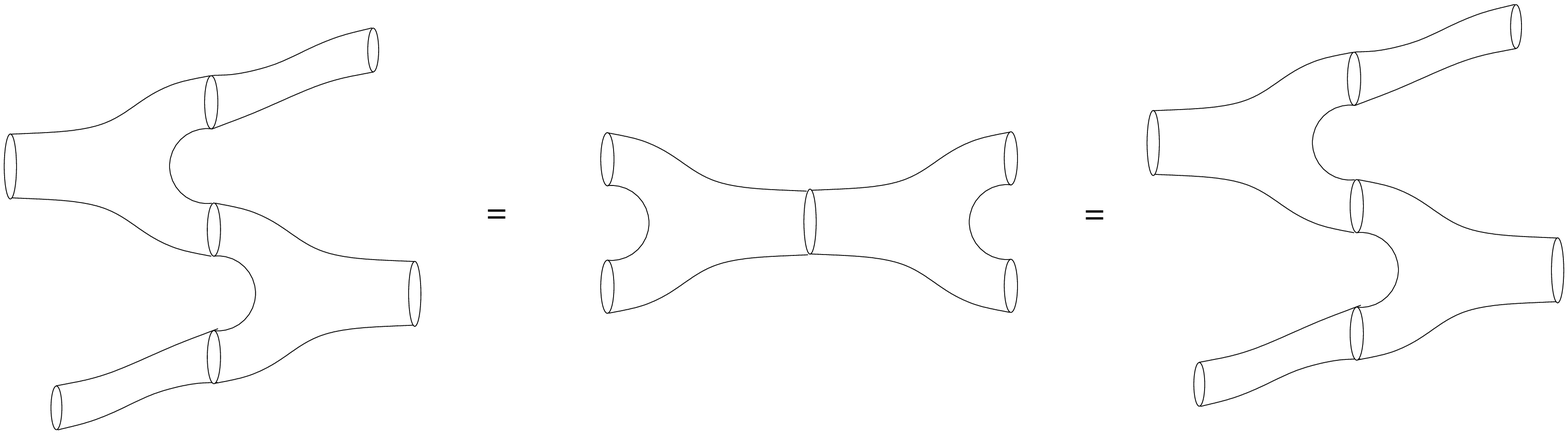}
  \caption{Frobenius relation}
\end{figure}

\begin{figure}[!h]
  \centering
  \includegraphics[width=1\textwidth]{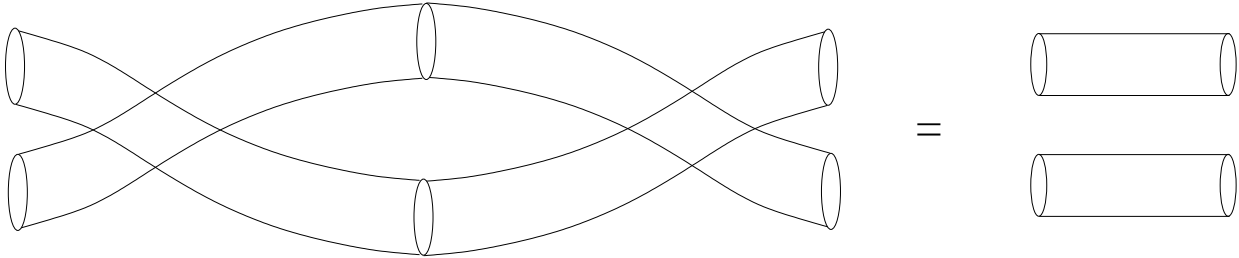}
  \caption{symmetry}
\end{figure}

\indent
    From the theorem and the definition of TQFT, we can construct
the following beautiful correspondence:\\
\indent {\bf Theorem 2.4.2} There is a canonical equivalence of
categories
\begin{equation}
2TQFT_k \simeq cFA_k
\end{equation}
where $2TQFT=Rep_k(2Cob)=SymMonCat(2Cob,Hilb_k)$ whose objects are
the symmetric monoidal functors from $2Cob$ to $Hilb_k$, and whose
arrows are the monoidal natural transformation between such
functors. $cFA_k$ denote the category of commutative Frobenuis
algebras over $k$ and Frobenuis algebra homomorphisms.

\subsection{$1-2-3$ theorem}
\indent
    The above description of $nd-TQFT$ always read "$(n-1)-n$" theory,
for example, the $0-1$ theory $Z$ assigns a vector space $Z(M)$ to
every closed 0-manifold $M$. A zero dimensional manifold $M$ is
simply a finite set of points. An orientation of $M$ determines a
decomposition $M=M_{+} \sqcup M_{-}$ of $M$ into "positively
oriented" and "negatively oriented" points. In particular, there are
two oriented manifolds which consist of only a single point, up to
orientation-preserving diffeomorphism. let us denote these manifolds
by $P$ and $Q$, then we obtain vector spaces $Z(P)$ and $Z(Q)$ where
$Z(Q)$ is dual space of $Z(P)$. We write $Z(P)=V$ and $Z(Q)=
\check{V}$, for some finite-dimensional vector space $V$. We must
also specify the behavior of $Z$ on $1$-manifolds $B$ with boundary
which is diffeomorphic either to a closed interval $[0,1]$ or to a
circle $S^{1}$. There are five cases to consider, depending on how
we decompose $\partial B$ into "incoming" and "outgoing" pieces.
\begin{itemize}
\item Suppose that $B=[0,1]$, regarded as a cobordism from $P$ to
itself. Then $Z(B)$ coincides with the identity map $id: V
\longrightarrow V$.
\item Suppose that $B=[0,1]$, regarded as a cobordism from $Q$ to
itself. Then $Z(B)$ coincides with the identity map $id: \check{V}
\longrightarrow \check{V}$.
\item Suppose that $B=[0,1]$, regarded as a cobordism from $P
\sqcup Q$ to the empty set. Then $Z(B)$ is a linear map from $V
\otimes \check{V}$ into ground field $k$: namely, the evaluation map
$(\upsilon,\lambda) \longmapsto \lambda{\upsilon}$.
\item Suppose that $B=[0,1]$, regarded as a cobordism from the empty set to
$P \sqcup Q$. Then $Z(B)$ is a linear map from $k$ to $V \otimes
\check{V}$. Under the canonical isomorphism $P \sqcup Q \simeq
End(V)$, this linear map is given by $x \longmapsto xid_V$.
\item Suppose that $B=S^1$, regarded as  a cobordism from the empty
s et to itself. Then $Z(B)$ is a linear map from $k$ to itself,
which we can identity with an element of $k$. Decomposing $S^{1}
\simeq {z\in C:|z|=1}=S_{-}^1 \cup S_{+}^1$ where $S_{-}^1={z \in C:
(|z|=1)\cap Im(z)\le 0}$ and $S_{+}^1={z \in C: (|z|=1)\cap Im(z)\ge
0}$ meeting in the subset $S_{-}^1 \cap S_{+}^1 ={\pm 1}$, then we
get the composition of the maps
\begin{equation}
k \simeq Z(\o) \longrightarrow Z(\pm 1) \longrightarrow Z(\o) \simeq
k
\end{equation}
Consequently, $Z(S^1)$ is given by the trace of the identity map
from $V$ to itself: in other words, the dimension of $V$.
\end{itemize}
\indent
    From the "$0-1$" theory, we can get the invariants via cutting and
gluing the cobordism, and an element of $k$ to each closed
$n$-manifold and a vector space to any closed $(n-1)$-manifold. What
we will gain to $(n-2)$-manifold? This just concerns with Extend
TQFT which is also called $0-1-2- \cdots -n$ theory. There is a very
beautiful correspondence between $0-1-2- \cdots -n$ theories and
dualizable objects [BD][L].\\
\indent {\bf Theorem }(Baez-Dolan-Lurie) Let $\cal{C}$ be a symmetric
monoidal infinity $n$-category. Then the space of $0-1-2- \cdots -n$
theories of framed manifolds with values in $\cal{C}$ is homotopy
equivalent to the space of fully dualizable objects in $\cal{C}$. \\
\indent
   In this paper, we don't refer this topic. We will see what
does Chern-Simons with compact Lie group attach to a point\ (i.e.
$0-1-2-3$
theory)\ in part 2 via this theorem.\\
\indent
    For $1-2-3$ theory, we have the following theory [BK] \\
\indent
  {\bf Theorem} A $1-2-3$ theory $F$ determines a modular tensor
category ${\cal C}= F(S^1)$. Conversely, a modular tensor category
$\cal{C}$
determines a $1-2-3$ theory $F$ with $ F(S^1)=\cal{C}$. \\
\indent
   I do not know a structure theorem for $2-3$ theories (i.e. $3d$-TQFT) like the
theorem 2.4.2. There is an open problem: can $3d- TQFTs$ distinguish
closed 3d manifold? It is a very interesting problem which I concern
all the time.

\section{Chern-Simons Theory}
\indent
   In this section, we will focus on the Chern-Simons theory [F2].\\
\indent
   Let $X$ be a $3d$-manifold and $G$ be a connected, simply connected,
compact Lie group, and an invariant form $\langle \rangle$ on its
Lie algebra $g$. Define the category ${\cal C}_X ={\cal C}_X^G$ of
$G$ connections as follows. An object in ${\cal{C}}_X$ is a
connection $\Theta$ on a principal $G$ bundle $P \longrightarrow X$.
A morphism $\Theta^{'} \longrightarrow \Theta$ is a bundle map
$\psi: P^{'} \longrightarrow P$ covering the identity map on $X$
(i.e. a bundle morphism) such that $\Theta^{'} =\psi^{\ast}\Theta$.
Obviously, such a category is a groupoid. Let $A_P$ of all
connections on principal $G$ bundle $P\longrightarrow X$, it is an
affine subspace of $\Omega_{P}^1(g)$, the vector space of $g$-valued
$1$-forms on $P$. Then the objects of $C_X$ form a union of affine
space
\begin{equation}
Obj(C_X)=\sqcup A_P \nonumber
\end{equation}
where ${P}$ is the collection of all principal $G$ bundles over
$X$.\\
\indent
    Before we give the theorem of Chern-Simons action, we first make
the following integrating hypothesis on the bilinear form $\langle
\rangle$.\\
\indent {\bf Hypothesis} Assume that the closed form
$-\frac16\langle \theta \wedge [\theta \wedge \theta]\rangle$
represents an integral class in $H^3(G;R)$ ,\
where $\theta$ is the Maurer-Cartan form.\\
\indent
    Fix a $G$ connection $\eta$ on $Q \longrightarrow Y$, where $Y$
is an oriented closed $2$-manifold. Let $\cal{C}_Q$ be the category
whose objects are sections $q: Y\longrightarrow Q$. For any two
sections $q, q^{'}$ there is a unique morphism $\psi:
q\longrightarrow q^{'}$, where $\psi :Q\longrightarrow Q$ is the
gauge transformation such that $q^{'} =\psi q$. Define the functor
$\Xi_{\eta} :{\cal C}_Q \longrightarrow \L$ by
$\Xi_{\eta}(q)={\mathbb C}$ for all $q$, where $\mathbb C$ has its
standard metric, and $\Xi_{\psi}:q \longrightarrow q^{'}$ is a
multiplication by $c_{Y}(q^{\ast}\eta,g_{\psi} q)$, where
$g_{\psi}:Q \longrightarrow G$ is the map associated to $\psi$, and
$c_{Y}$ is the cocycle
\begin{equation}
c_{Y}(a,g)=exp(2{\pi}i \int_Y(\langle Ad_{g^{-1}}a\wedge
\phi_g\rangle +W_{Y}(g)),   a\in
\Omega_{Y}^1(\frak{g}),\quad g:Y\longrightarrow G
\end{equation}
\indent
   We obtain the metrized line $L_{\eta}=L_{Y,\eta}$ of
invariant section.\\
\indent
   We now give the theorem of Chern-Simons action
$S_{X}(p,\theta)=\int_X(p_{\ast}\alpha(\Theta)$, where
$p:X\longrightarrow P$ be a section and $\alpha(\Theta)$ be the
Chern-Simons form as the lagrangian of field $\Theta$.\\
\indent
    {\bf Theorem 3}\ Let $G$ be a connected, simply connected compact
Lie group and $\langle\rangle: \frak{g}\otimes \frak{g}
\longrightarrow {\mathbb R} $ an invariant form on its Lie algebra
$g$ which satisfies the integrality condition Hypothesis. Then the
assignments
\begin{eqnarray}
&&\eta \mapsto L_{\eta} ,\qquad \eta \in {\cal C}_{Y} \\
&&\Theta \mapsto e^{2 \pi i S_{X}(\Theta)}, \qquad \Theta
\in {\cal C}_{X}
\end{eqnarray}
defines above for closed oriented 2-manifolds $Y$ and compact
oriented $3$-manifolds $X$ are smooth and satisfy:
\begin{itemize}
\item (Functoriality) If $\psi: Q^{'} \longrightarrow Q$ is a bundle
map covering an orientation preserving difeomorphism $\bar{\psi}:
Y^" \longrightarrow Y$, and $\psi$ is a connection on $Q$, then
there is an induced isometry
\begin{equation}
\psi^{\ast} : L_{\eta} \longrightarrow L_{\psi^{\ast}\eta}
\end{equation}
and these compose properly. If $\varphi : P^{'} \longrightarrow P$
is a bundle map covering an orientation preserving diffeomorphism
$\bar{\varphi} : X^{'}\longrightarrow X$, and $\Theta$ is a
connection on $P$, then
\begin{equation}
{\partial \varphi}^{\ast}{e^{2\pi iS_{X}{\Theta}}} =e^{2\pi i
S_{X^{'}}{\varphi^{\ast}\Theta}},
\end{equation}
where $\partial \varphi: \partial P^{'} \longrightarrow \partial P$ is
the induced map over the boundary.
\item (Orientation) There is a natural isometry
\begin{eqnarray}
&&L_{-Y,\eta} \cong \overline{L_{Y,\eta}},\\
&&e^{2 \pi S_{-X}(\Theta)} =\overline{e^{2 \pi S_{X}(\Theta)}}
\end{eqnarray}
\item (Additivity) If $Y=Y_1 \sqcup Y_2$ is a disjoint union, and
$\eta_i$ are connections over $Y_i$, then there is a natural
isometry
\begin{equation}
L_{\eta_1 \sqcup \eta_2} \cong L_{\eta_1} \otimes L_{\eta_2}.
\end{equation}
If $X=X_1 \sqcup X_2$ is a disjoint union, and $\Theta_i$ are
connections over $X_i$, then
\begin{equation}
e^{2\pi i S_{X_1 \sqcup X_2}(\Theta_1 \sqcup \Theta_2)}=e^{2\pi i
S_{X_1}(\Theta_1)} \otimes e^{2\pi iS_{X_2}(\Theta_2)}.
\end{equation}
\item (Gluing) Suppose $Y\hookrightarrow X$ is a closed, oriented
submanifold and $X^{cut}$ is the manifold obtained by cutting $X$
along $Y$. Then $\partial X^{cut}=\partial X \sqcup Y\sqcup (-Y)$.
Suppose $\Theta$ is a connection over $X$, with $\Theta^{cut}$ the
induced connection over $X^{cut}$, and $\eta$ the restriction of
$\Theta$ to $Y$. Then
\begin{equation}
e^{2\pi i S_{X}(\Theta)}={Tr}_{\eta}(e^{2\pi i S_{X^{cut}}(\Theta^{cut})})
\end{equation}
where ${Tr}_{\eta}$ is the contraction
\begin{equation}
{Tr}_{\eta} : L_{\partial{\Theta^{cut}}} \otimes L_{\eta} \otimes
\overline{L_{\eta}} \longrightarrow L_{\partial \theta}
\end{equation}
\end{itemize}

\indent
   We usually use the "gluing" law to compute the invariants of
manifold, and the "additivity" law is crucial point to construct the
TQFT via Chern-Simons theory.\\

\section{Span and Cospan}
\indent
    To construct the TQFT via Chern-simons theory, we firt need to construct
 a functor $\natural$ from $nCob$ to $Span(FT)$ of field space category.
 Let us give the definition of Span and Cospan. \\

\subsection{ The category Span(FT)}
\indent
    {\bf Definition 4.1} \ Given any category $\cal{C}$, a span
$(S,s,t)$ between objects $ X_1 ,X_2 \in \cal{C}$ is a diagram in
$\cal{C}$ of the form
\begin{equation}
X_1 \longleftarrow S \longrightarrow X_2 \nonumber
\end{equation}
\indent
    {\bf Definition 4.2} \ Given two spans $(S,s,t)$ and $(S^{'}, s^{'},t^{'})$ between $X_1$ and
$X_2$, a morphism of spans is a morphism $g: S \longrightarrow S^{'}$
making the diagram commutes. \\
\indent
    Composition of spans $S$ from $X_1$ to $X_2$ and $S^{'}$ from
$X_2$ to $X_3$ is given by pullback: that is, an object $R$ with
maps $f_1$ and $f_2$ making the diagram which satisfies the "universal property" commutes.\\

 \indent
    {\bf Definition 4.3} \ A cospan in $\cal{C}$ is a span in ${\cal{C}}^{op}$,
morphisms of cospan are morphisms of span in ${\cal{C}}^{op}$, and
composition of cospans is given by pullback in ${\cal{C}}^{op}$-that
is, by pushout in $\cal{C}$.\\
\indent
     From the definitions of "Cobordism" and "Cospan", obviously, a
cobordism from $\Sigma_1$ to $\Sigma_2$ can viewed as a cospan from
$\Sigma_1$ to $\Sigma_2$.\\
\indent
     As we describe above, for a manifold $X$ with $\partial X = Y_0
\sqcup Y_1 \ (Y_0 {\rm\ and\ } Y_1 {\rm \ may\  be\ } \o)$, We have a groupoid
${\cal{C}}_{X}$ which is the field space of Chern-Simons theory. We
can construct ${\cal{C}}_{Y_0}$ and  ${\cal{C}}_{Y_1}$ by
restriction $\iota_0: {\cal{C}}_{X} \longrightarrow {\cal{C}}_{Y_0}$
and $\iota_1: {\cal{C}}_{X} \longrightarrow {\cal{C}}_{Y_1}$.
Obviously, $({\cal{C}}_{X},\iota_0,\iota_1)$ is a span between
${\cal C}_{Y_0}$ and ${\cal C}_{Y_1}$. Then we obtain the category
$Span(FT)$ which is a monoidal category, composition as its monoidal
product.

\subsection{Functor from 3Cob to Span(FT)}
\indent
     From the above consideration, we have the following property.\\
\indent
    {\bf Property 4.2} There is a monoidal functor \\
\begin{eqnarray}
\natural:&&3Cob \longrightarrow Span(FT) \\ \nonumber
&&Y \longmapsto {\cal C}_{Y}\\
&&Y_0 \longrightarrow X \longleftarrow Y_1 \longmapsto {\cal C}_{Y_0}
\longleftarrow {\cal C}_{X} \longrightarrow {\cal{C}}_{Y_1}
\end{eqnarray}
where $Y$ is an object of $3Cob$. \\
\indent
   {\bf Proof}.\ The check is straightforward from the above details.

\section {TQFT via Chern-Simons Theory}
\indent
   Now we begin to construct the TQFT. We need another monoidal
functor $\heartsuit$
\begin{equation}
\heartsuit : (Span(FT)) \longrightarrow {Hilb}
\end{equation}

\subsection{The monoidal functor $\heartsuit$}
\indent
    For any closed 2-manifold $Y$ associated to
${\cal{C}}_{Y}$, an object of $Span(FT)$, we can get a metrized line
$L_{\eta}$ to each $\eta \in {\cal{C}}_{Y}$. Assume there exist
measures ${\mu}_{X}$, ${\mu}_{Y}$ on the spaces ${\cal{C}}_{X}$,
${\cal{C}}_{Y}$, then define the Hilbert space
$H_{Y}=L^{2}({\cal{C}}_{Y})$. For the morphism ${\cal C}_{Y_0}
\longleftarrow {\cal C}_{X} \longrightarrow {\cal{C}}_{Y_1}$ of the
monoidal category $Span(FT)$, we define the linear map as the
push-pull-with-kernel $e^{iS_{X}}$:
\begin{equation}
t_{\ast} \circ e^{iS_{X}} \circ s^{\ast} : H_{Y_0} \longrightarrow
H_{Y_1}
\end{equation}
\indent
   In short,we get the monoidal functor as follow.\\
\indent
   {\bf Theorem 5.1}\ There is a monoidal functor $\heartsuit$ from
the category $Span(FT)$ to $Hilb$ assigning
\begin{eqnarray}
{\cal C}_{Y} &\longmapsto& H_{Y} \\
{\cal C}_{Y_0} \longleftarrow {\cal C}_{X} \longrightarrow
{\cal{C}}_{Y_1} &\longmapsto& t_{\ast} \circ e^{iS_{X}} \circ s^{\ast}
: H_{Y_0} \longrightarrow H_{Y_1}
\end{eqnarray}
where $t:{\cal C}_{X} \longrightarrow {\cal C}_{Y_0}$ and $s:{\cal
C}_{X} \longrightarrow {\cal C}_{Y_1}$.\\
 \indent
   {\bf Proof}. \ From the Chern-Simons theory, we can easily to get ${\cal{C}}_{Y} \longmapsto
H_{Y}$,
and the above constructions just interpret that $\heartsuit$
is a functor. What we need to check is that this functor is a
monoidal, i.e.
\begin{equation}
{t^{'}}_{\ast} \circ \exp(iS_{X^{'}}) \circ {s^{'}}^{\ast} \circ
t_{\ast} \circ \exp(iS_{X}) \circ s^{\ast} ={t^{'}}_{\ast} \circ
{r^{'}}_{\ast} \circ \exp(iS_{X^{'}\circ X}) \circ r^{\ast} \circ
s^{\ast}
\end{equation}
where $t^{'} :{\cal{C}}_{X^{'}}\longrightarrow {\cal{C}}_{Y_2},\
s^{'} : {\cal{C}}_{X^{'}}\longrightarrow {\cal{C}}_{Y_1} ,\  r :
{\cal{C}}_{{X^{'}} \circ {X}} \longrightarrow {\cal{C}}_{X}, \rm \
and \ r^{'} :{\cal{C}}_{{X^{'}} \circ {X}} \longrightarrow
{\cal{C}}_{X^{'}},$\\
 and
\begin{equation}
\heartsuit ({\cal{C}}_{Y_0 \circ Y_1})= H_{Y_0} \otimes H_{Y_1}.
\end{equation}
\indent
 (25)is hold, since the pushforward $t_{\ast}$ and
${t^{'}}_{\ast}$ are integrations and $e^{2\pi i S_{X \sqcup
X^{'}}(\Theta_1 \sqcup \Theta_2)}=e^{2\pi i S_{X}(\Theta_1)} \otimes
e^{2\pi iS_{X^{'}}(\Theta_2)}$.\\
\indent
 And (26) is hold, since $L_{\eta_1 \sqcup \eta_2} \cong
L_{\eta_1}
\otimes L_{\eta_2}$. \\
\indent
The proof is complete.\\
\indent
   From the above proof,we can conclude that the functor
$\heartsuit$ is equivalent to the "additivity" law of Chern-Simons
theory.

\subsection{Construction of TQFT}
\indent
   Using the above two monoidal functors, we can construct our
$3d$-TQFT.\\
\indent
  {\bf Theorem 5.2} The functor
\begin{equation}
Z=\heartsuit \circ \natural
\end{equation}
is a $3d$-TQFT.\\
 \indent
   {\bf Proof}.\ Since both $\heartsuit$ and $\natural$ are monoidal
functors, so is $Z$. Then $Z$ is a $3d$-TQFT from the definition of
$nd$-TQFT. The proof is complete.\\

\section{Conclusion}
\indent
    For abstract TQFT, how to construct a physical one is very
interesting problem. In this paper, we construct a TQFT via
Chern-Simons theory, which provides invariants of 3d-manifolds,
though it appeared in Freed's paper as axioms. In fact, ideas of
categorification has been in physical theory [BL], which also can be
viewed as our main philosophy: how "high-algebraic" ideas from
category theory can illuminate questions in string theory , quantum
field theory and geometry.\\
 \indent
   In the next part, we will see how the higher structure function
in the "Extended TQFT", which ultimately concern with the
Chern-Simons theory attach to point.

\section{ Acknowledgements}
\indent
    First and foremost, I would like to thank the group directed by Professor Ke Wu (my
advisor) and Professor Shikun Wang at Morningside Center of
Mathematics(MCM ) for giving me the opportunity to take part in and
give my topics in the group. I learned much from the group and
discussed freely with the members in the group. Secondly, I would
like to sincerely thank Dr. Jie Yang for her endless assistance.\\

\end{document}